\documentclass
[superscriptaddress,secnumarabic,amssymb,amsmath,nobibnotes,aps,prd,showkeys,showpacs,nofootinbib,onecolumn,12pt]{revtex4}%
\usepackage{graphicx}
\usepackage{subfigure}
\usepackage{epsf}
\usepackage{bm}
\usepackage{amsmath}
\usepackage{amsfonts}
\usepackage{amssymb}%
\providecommand{\U}[1]{\protect\rule{.1in}{.1in}}

\newcommand{\be}{\begin{equation}}
\newcommand{\ee}{\end{equation}}

\newcommand{\mincir}{\raise
-3.truept\hbox{\rlap{\hbox{$\sim$}}\raise4.truept\hbox{$<$}\ }}
\newcommand{\magcir}{\raise
-3.truept\hbox{\rlap{\hbox{$\sim$}}\raise4.truept\hbox{$>$}\ }}

\begin{document}
\title{Painlev\'{e} analysis for the cosmological field equations in Weyl Integrable Spacetime}
\author{Andronikos Paliathanasis}
\email{anpaliat@phys.uoa.gr}
\affiliation{Institute of Systems Science, Durban University of Technology, Durban 4000,
South Africa}
\affiliation{Instituto de Ciencias F\'{\i}sicas y Matem\'{a}ticas, Universidad Austral de
Chile, Valdivia 5090000, Chile}

\begin{abstract}
We apply singularity analysis to investigate the integrability properties of
the gravitational field equations in Weyl Integrable Spacetime for a spatially
flat Friedmann--Lema\^{\i}tre--Robertson--Walker background spacetime induced
with an ideal gas. We find that the field equations possess the Painlev\'{e}
property in the presence of the cosmological constant and the analytic
solution is given by a left Laurent expansion.

\end{abstract}
\date{\today}
\maketitle

\section{Introduction}

\label{sec1}

The Weyl Integrable Spacetime (WIS) is a natural way to extend Einstein's
General Relativity, in which a scalar field is introduced in the natural space
by geometrical degrees of freedom \cite{salim96}. Scalar fields play an
important role in the description of gravitational phenomena at large scales
\cite{obs1,obs2}. Indeed, the late time or early time acceleration phases of
the universe have been proposed to be attributed to scalar fields
\cite{sf1,sf2,sf3,sf5,sf6}. Another novelty of WIS is that interaction is
introduced geometrically between the matter components \cite{ch1,ch2,ch3}.
Cosmological models with interaction in the dark sector of the universe have
been widely studied before. Furthermore, that interacting models survive the
constraints follow by the cosmological observations \cite{os1,os2,os3}.
Specifically, the interaction in the dark sector of the universe is a
mechanism to solve the cosmic coincidence problem and it has been used for the
explanation of the discrepancy in the cosmological constant
\cite{Amendola-ide1,Amendola-ide2,Pavon:2005yx,delCampo:2008jx,Wetterich-ide1}%
. In addition, WIS is related with the chameleon cosmology \cite{cham1}.

There are various studies in the literature on cosmological models on WIS.
Multi-dimensional spacetimes were investigated in \cite{wm1}, while
lower-dimensional gravitational models were studied in\cite{wm1a}. Inflation
is WIS investigated in \cite{wm2,wm3}, where it was found that nonsingular
inflationary solutions are provided by the specific theory.\ In
\cite{wm4,ww4,ww5} WIS has been considered as the dark energy mechanism which
drives the present acceleration of the universe, while recently an extension
of WIS in Einstein-Aether theory was studied in \cite{ww3}. The dynamical
analysis of the field equations with different fluid sources for the
construction of the cosmological history was the subject of study in
\cite{wm6}.

The field equations in WIS are nonlinear ordinary differential equations of
second-order. Because of the new degrees of freedom provided by the scalar
field there are not many known solutions in the literature. Note that the
cosmological field equations in WIS admit a minisuperspace description the
Noether symmetry analysis for which was applied in \cite{wm7} for the
construction of conservation laws and the derivation of analytic solutions.
The Noether symmetry analysis is a powerful and systematic approach for the
study of the integrability properties of dynamical systems which follow from a
variational principle with many applications in cosmological studies
\cite{ns1,ns2,ns3,ns4,ns5,ns6}. For the cosmological model in WIS in which the
matter source is an ideal gas it was found that the introduction of the
cosmological constant term into the gravitational Action Integral leads to a
dynamical system with fewer Noetherian conservation laws from which we are not
able to infer the integrability properties. In this work we investigate the
same problem with the use of another important tool of analysis for the
construction of conservation laws and analytic solutions, known as
Painlev\'{e} analysis or singularity analysis \cite{buntis}. The singularity
analysis for the analysis of the field equations in gravity covers a wide
range of applications; see for instance
\ \cite{ftAn,ksAn,cots11,cots2,Feix97,pcosm,bun1,cots}. The determination of
the integrability properties of dynamical systems is essential in physics and
in all areas of applied mathematics. The novelty when a physical system is
described by an integrable dynamical system is that we know that an actual
solution exists when we apply numerical methods for the study of the system,
or there exists a closed-form function which solves the dynamical system. For
a recent discussion on the integrable cosmological models we refer the reader
to \cite{tw1}. In our study the singularity analysis leads to the construction
of analytic solutions expressed as Laurent expansions around a movable
singularity. The plan of the paper is as follows.

In Section \ref{sec2} we present the cosmological model of our consideration
which is that of WIS in an homogeneous and isotropic universe in which the
matter source is that of an ideal gas. In Section \ref{sec3} we review
previous results wherein we show the existence of additional conservation laws
for the field equations in the absence of the cosmological constant term.
Section \ref{sec4} includes the main analysis of this study, in which the
singularity analysis is applied for the study of the integrability properties
for the field equations. We find that the field equations possess the
Painlev\'{e} property with or without a cosmological constant term and the
analytic solution is expressed by Laurent expansions. Finally, in Section
\ref{sec5} we summarize our results and we draw our conclusions.

\section{Field equations{}}

\label{sec2}

In WIS or Weyl Integrable Geometry (WIG), the modified Einstein-Hilbert Action
is expressed as
\begin{equation}
S_{W}=\int dx^{4}\sqrt{-g}\left(  \tilde{R}+\xi\left(  \tilde{\nabla}_{\nu
}\left(  \tilde{\nabla}_{\mu}\phi\right)  \right)  g^{\mu\nu}-\Lambda
+\mathcal{L}_{m}\right)  ,
\end{equation}
in which $g_{\mu\nu}$ is the metric tensor for the physical space,
$\tilde{\nabla}_{\mu}$ denotes covariant derivative defined by the symbols
$\tilde{\Gamma}_{\mu\nu}^{\kappa}$, where $\tilde{\Gamma}_{\mu\nu}^{\kappa}$
are the Christoffel symbols for the conformally related metric $\tilde{g}%
_{\mu\nu}=\phi g_{\mu\nu}$. The scalar field $\phi$ is the coupling function,
$\Lambda$ is the cosmological constant term, the parameter $\xi$ is an
arbitrary coupling constant and $\mathcal{L}_{m}$ is the Lagrangian function
for the matter source.

When $\mathcal{L}_{m}$ describes a perfect fluid with energy density $\rho$
and pressure component $p$, the field equations in the Einstein-Weyl theory
are derived to be%
\begin{equation}
\tilde{G}_{\mu\nu}+\tilde{\nabla}_{\nu}\left(  \tilde{\nabla}_{\mu}%
\phi\right)  -\left(  2\xi-1\right)  \left(  \tilde{\nabla}_{\mu}\phi\right)
\left(  \tilde{\nabla}_{\nu}\phi\right)  +\xi g_{\mu\nu}g^{\kappa\lambda
}\left(  \tilde{\nabla}_{\kappa}\phi\right)  \left(  \tilde{\nabla}_{\lambda
}\phi\right)  -\Lambda g_{\mu\nu}=-\left(  \tilde{\rho}+\tilde{p}\right)
u_{\mu}u_{\nu}-\tilde{p}g_{\mu\nu}, \label{ww.08}%
\end{equation}
where $\tilde{G}_{\mu\nu}$ is the Einstein tensor with respect the metric
$\tilde{g}_{\mu\nu}$ and $u^{\mu}$ is the comoving observer. Parameters
$\tilde{\rho},~\tilde{p}\,\ $are the energy density and pressure~components
for the matter source multiplied by the factor $e^{-\frac{\phi}{2}}$

The field equations (\ref{ww.08}) can be written in the equivalent form
\begin{equation}
G_{\mu\nu}-\lambda\left(  \phi_{,\mu}\phi_{,\nu}-\frac{1}{2}g_{\mu\nu}%
\phi^{,\kappa}\phi_{,\kappa}\right)  -\Lambda g_{\mu\nu}=-\left(  \tilde{\rho
}+\tilde{p}\right)  u_{\mu}u_{\nu}-\tilde{p}g_{\mu\nu}, \label{ww.11}%
\end{equation}
where $G_{\mu\nu}$ is the Einstein tensor for the background space $g_{\mu\nu
}$ and $\lambda=2\xi-\frac{3}{2}$.

For the homogeneous and isotropic spatially flat
Friedmann--Lema\^{\i}tre--Robertson--Walker (FLRW) spacetime,
\begin{equation}
ds^{2}=-N^{2}\left(  t\right)  dt^{2}+a^{2}\left(  t\right)  \left(
dr^{2}+r^{2}\left(  d\theta^{2}+\sin^{2}\theta d\varphi^{2}\right)  \right)
,\label{ww.16}%
\end{equation}
the modified Friedmann's equations are as follow%
\begin{equation}
3H^{2}-\frac{\lambda}{2N^{2}}\dot{\phi}^{2}-\Lambda-e^{-\frac{\phi}{2}}%
\rho=0,\label{ww.17}%
\end{equation}%
\begin{equation}
\dot{H}+H^{2}+\frac{1}{6}e^{-\frac{\phi}{2}}\left(  \rho+3p\right)
+\frac{\lambda}{3N^{2}}\dot{\phi}^{2}-\frac{\Lambda}{3}=0,\label{ww.18}%
\end{equation}%
\begin{equation}
\ddot{\phi}-\frac{\dot{N}}{N}\dot{\phi}+3H\dot{\phi}+\frac{1}{2\lambda}%
N^{2}e^{-\frac{\phi}{2}}\rho=0\label{ww.19a}%
\end{equation}
and
\begin{equation}
\dot{\rho}+3NH\left(  \rho+p\right)  -\rho\dot{\phi}=0\label{ww.20a}%
\end{equation}
in which we have considered the comoving observer, $u^{\mu}=\frac{1}{N}%
\delta_{t}^{\mu}$, and $H=\frac{1}{N}\frac{\dot{a}}{a}$ is the Hubble function.

In this study we consider that the matter source is that of an ideal gas, that
is $p=\left(  \gamma-1\right)  \rho,~\gamma<2$. For $\gamma=2$ a two scalar
field model is recovered \cite{tw2}.From equation (\ref{ww.20a}) it follows
that $\rho=\rho_{0}a^{-3\gamma}e^{\phi}$. Thus we end with the set of
differential equations (\ref{ww.17}), (\ref{ww.18}) and (\ref{ww.19a}).

The cosmological history for this specific cosmological model investigated
before in \cite{wm6}. From the analysis of the dynamics it was found that the
cosmological history is consisted by two matter epochs and two acceleration
phases. The matter epochs correspond to unstable asymptotic solutions while
for the two accelerated phases the one is always unstable which can be
corresponded to the early acceleration phase of the universe, and the second
accelerated asymptotic solution describes the future de Sitter attractor.

\section{Lagrangian description}

\label{sec3}

In the case of an ideal gas, the cosmological field equations (\ref{ww.17}),
(\ref{ww.18}) and (\ref{ww.19a}) can be reproduced by the variation of the
point-like Lagrangian
\begin{equation}
\mathcal{L}\left(  N,a,\dot{a},\phi,\dot{\phi}\right)  =\frac{1}{N}\left(
-3a\dot{a}^{2}+\frac{\lambda}{2}a^{3}\dot{\phi}^{2}\right)  -N\left(
a^{3}\Lambda+\rho_{m0}e^{\frac{\phi}{2}}a^{3-3\gamma}\right)  . \label{ee.04}%
\end{equation}

The Lagrangian function (\ref{ee.04}) is a singular Lagrangian because
$\frac{\partial\mathcal{L}}{\partial\dot{N}}=0.$ The constraint equation
(\ref{ww.17}) is the Euler-Lagrange equation with respect to the lapse
function $N\left(  t\right)  $, i.e. $\frac{\partial\mathcal{L}}{\partial
N}=0$, while the second-order differential equations, (\ref{ww.18}) and
(\ref{ww.19a}), are the Euler-Lagrange equations with respect to the variables
$a\left(  t\right)  $ and $\phi\left(  t\right)  $ respectively, i.e.
$\frac{d}{dt}\left(  \frac{\partial\mathcal{L}}{\partial\dot{a}}\right)
-\frac{\partial\mathcal{L}}{\partial a}=0$ ; $\frac{d}{dt}\left(
\frac{\partial\mathcal{L}}{\partial\dot{\phi}}\right)  -\frac{\partial
\mathcal{L}}{\partial\phi}=0$.

The existence of the singular Lagrangian (\ref{ee.04}) and the minisuperspace
description of the gravitational model are important characteristics for the
study of the integrability properties of the field equations. Without loss of
generality the lapse function can be assumed to be $N\left(  t\right)
=N\left(  a\left(  t\right)  ,\phi\left(  t\right)  \right)  $. In such a
consideration, the Lagrangian (\ref{ee.04}) is regular while equation
(\ref{ww.17}) is the conservation law of \textquotedblleft
energy\textquotedblright, that is, the Hamiltonian of the dynamical system.

\subsection{Normal coordinates}

We follow \cite{wm6} and without loss of generality we select $N\left(
t\right)  =a\left(  t\right)  ^{3}$. Furthermore, we define the new field
$\Phi$,
\begin{equation}
\phi=\frac{1}{\lambda\left(  2-\gamma\right)  }\left(  2\lambda\ln\Phi-\ln
a\right)  . \label{ee.05}%
\end{equation}
In the new variables $\left\{  a,\Phi\right\}  $ the Lagrangian of the field
equations is%
\begin{equation}
\mathcal{L}\left(  a,\dot{a},\Phi,\dot{\Phi}\right)  =\frac{1}{2}\left(
6-\frac{1}{\left(  \gamma-2\right)  ^{2}\lambda}\right)  \left(  \frac{\dot
{a}}{a}\right)  ^{2}+\frac{2}{\left(  \gamma-2\right)  ^{2}}\frac{\dot{a}\Phi
}{a\Phi}-\frac{2\lambda}{\left(  \gamma-2\right)  ^{2}}\left(  \frac{\dot
{\Phi}}{\Phi}\right)  ^{2}+\Lambda a^{6}+a^{3\left(  2-\gamma\right)
-\frac{1}{2\lambda\left(  2-\gamma\right)  }}\Phi^{\frac{1}{2-\gamma}}%
\text{~}. \label{ee.06}%
\end{equation}
The field equations are%
\begin{equation}
\frac{1}{2}\left(  6-\frac{1}{\left(  \gamma-2\right)  ^{2}\lambda}\right)
\left(  \frac{\dot{a}}{a}\right)  ^{2}+\frac{2}{\left(  \gamma-2\right)  ^{2}%
}\frac{\dot{a}\Phi}{a\Phi}-\frac{2\lambda}{\left(  \gamma-2\right)  ^{2}%
}\left(  \frac{\dot{\Phi}}{\Phi}\right)  ^{2}-\Lambda a^{6}-a^{3\left(
2-\gamma\right)  -\frac{1}{2\lambda\left(  2-\gamma\right)  }}\Phi^{\frac
{1}{2-\gamma}}=0, \label{ee.07}%
\end{equation}%
\begin{equation}
\ddot{a}+\frac{1}{2}a^{7-3\gamma-\frac{1}{2\lambda\left(  2-\gamma\right)  }%
}\Phi^{\frac{1}{2-\gamma}}-\frac{\dot{a}^{2}}{a}-\Lambda a^{7}=0 \label{ee.08}%
\end{equation}
and
\begin{equation}
\ddot{\Phi}-\frac{\dot{\Phi}^{2}}{\Phi}-\frac{\Lambda}{2\lambda}a^{6}\Phi=0.
\label{ee.09}%
\end{equation}

When $\Lambda=0$, equation (\ref{ee.09}) provides the conservation law
$I_{0}=\frac{d}{dt}\ln\Phi$. This conservation law was derived before in
\cite{wm6,wm7}. The set of variables $\left\{  a,\Phi\right\}  $ constitutes
the normal coordinates for the field equations. However, in the presence of
the cosmological constant the conservation law does not exist. The existence
of the second conservation law for the field equations is essential in order
to be able to conclude about the integrability of the field equations. Indeed,
in \cite{wm7} the Hamilton-Jacobi equation was solved and the field equations
were reduced into a system of two first-order ordinary differential equation.

In the presence of $\Lambda$, we applied the symmetry analysis for point and
contact symmetries and we find that the field equations do not admit any
Noether symmetry. Furthermore, we considered polynomial functions of $I\left(
a,\dot{a},\Phi,\dot{\Phi}\right)  $ to be a conservation law. However, we were
not able to determine any function $I$ with this specific requirement.

The concept of integrability is not limited in the existence of invariant
functions and conservation laws. According to the Painlev\'{e} approach, a
dynamical system is integrable if admits a movable pole and its solution is
expressed in terms of a Laurent expansion around the movable pole. In the
following we apply the singularity analysis in order to investigate the
integrability properties for the dynamical system of our consideration.

\section{Singularity analysis}

\label{sec4}

The modern treatment of singularity analysis is described by the ARS
algorithm. The algorithm has three main steps. They are (a) derivation of the
leading-order behaviour, (b) derivation of the resonances and (c) the
consistency test . For more details and examples on the application of the ARS
algorithm we refer the reader to \cite{buntis}.\ In the first step of the
algorithm we should show that there exists a movable singularity for the
dynamical system at which the solution is approximated by a singular
expression. For instance, for power-law expressions $\left(  \tau-\tau
_{0}\right)  ^{p}$, the exponent $p$ should be negative number. However, it
has been shown that $p$ can be also a rational number. The resonances should
be the same in number as the degrees of freedom of the problem and one of the
resonances has to be $-1$, while the resonances should be rational numbers. If
the requirements of the steps (a) and (b) are satisfied, we write the solution
in a Laurent expansion, in our case in Puiseux series, and we replace in the
original equation to check if it is an actual solution for the original
dynamical system.

We follow \cite{cots1} and we apply the singularity analysis for the
equivalent dynamical system in the dimensionless variables. Indeed, we follow
the $H-$normalization approach and we define the new variables%
\begin{equation}
x=\sqrt{\frac{1}{6}}\frac{\dot{\phi}}{H}~,~\Omega_{\Lambda}=\frac{\Lambda
}{3H^{2}}~,~\Omega_{m}=\frac{\rho_{m}}{3H^{2}}e^{-\frac{\phi}{2}}%
~,~d\tau=NHdt. \label{ww.25A}%
\end{equation}
where the latter parameter $\tau$ is the new independent variable.

Consequently, the field equations (\ref{ww.17})-(\ref{ww.20a}) for the ideal
gas are written as the following equivalent algebraic-differential system%
\begin{equation}
\frac{d\Omega_{\Lambda}}{d\tau}=-\Omega_{\Lambda}\left(  \left(
\gamma-2\right)  \lambda x^{2}+\gamma\left(  \Omega_{\Lambda}-1\right)
\right)  ~~ \label{ww.36A}%
\end{equation}
and
\begin{equation}
\frac{dx}{d\tau}=\frac{1}{12\lambda}\left(  \left(  \lambda x^{2}-1\right)
\left(  \sqrt{6}-6\left(  \gamma-2\right)  \lambda x\right)  +\left(  \sqrt
{6}-6\gamma\lambda x\right)  \Omega_{\Lambda}\right)  ~ \label{ww.37A}%
\end{equation}
with constraint equation%
\begin{equation}
\Omega_{m}=1-\lambda x^{2}-\Omega_{\Lambda}.
\end{equation}

Therefore we continue with the investigation of the singularity analysis for
the system of first-order ordinary differential equations (\ref{ww.36A}) and
(\ref{ww.37A}).

\subsection{Case $\Lambda=0$}

We focus now on the case $\Lambda=0$. Equation (\ref{ww.37A}) reduces to the
simple form%
\begin{equation}
\frac{dx}{d\tau}=\frac{1}{12\lambda}\left(  \left(  \lambda x^{2}-1\right)
\left(  \sqrt{6}-6\left(  \gamma-2\right)  \lambda x\right)  \right)  ~.
\label{ww.38A}%
\end{equation}
The closed-form solution of the this equation is
\begin{equation}
\tau-\tau_{0}=\frac{6\lambda\left(  \gamma-2\right)  }{1-6\lambda\left(
\gamma-2\right)  ^{2}}\ln\frac{\left(  \lambda x^{2}-1\right)  }{\left(
\sqrt{6}-6\lambda\left(  \gamma-2\right)  x\right)  ^{2}}-\frac{2\sqrt
{\lambda}}{1-6\lambda\left(  \gamma-2\right)  ^{2}}\arctan h\left(  \lambda
x\right)  .
\end{equation}
However, we are interested to investigate if equation (\ref{ww.38A}) possesses
the Painlev\'{e} property and if the analytic solution can be expressed in a
Laurent expansion around a movable pole.

According to the ARS algorithm, we replace $x\left(  t\right)  =x_{0}\left(
\tau-\tau_{0}\right)  ^{p}$ in (\ref{ww.38A}), that is,%
\begin{equation}
px_{0}\left(  \tau-\tau_{0}\right)  ^{-1+p}-\frac{1}{12\lambda}\left(  \left(
\lambda x_{0}^{2}\left(  \tau-\tau_{0}\right)  ^{2p}-1\right)  \left(
\sqrt{6}-6\left(  \gamma-2\right)  \lambda x_{0}\left(  \tau-\tau_{0}\right)
^{p}\right)  \right)  .
\end{equation}
Hence, the dominant terms provides the algebraic equation $-1+p=3p$, that is,
$p=-\frac{1}{2}$, and $\lambda\left(  \gamma-2\right)  x_{0}^{2}-1=0$.

For the determination of the resonances we substitute
\begin{equation}
x\left(  t\right)  =x_{0}\left(  \tau-\tau_{0}\right)  ^{-\frac{1}{2}%
}+m\left(  \tau-\tau_{0}\right)  ^{-\frac{1}{2}+S}~,~\lambda x_{0}^{2}+1=0,
\end{equation}
into (\ref{ww.38A}) and we expand around $m^{2}\rightarrow0$. Hence, we end
with the algebraic equation~$S+1=0$, which means that $S=-1$. \ That is in
agreement with the singularity analysis, because one of the resonances should
be $-1$. This resonance indicates that the leading-order behaviour describes a
movable singularity and $t_{0}$ is an integration constant.

For the third step of the ARS algorithm, known as the consistency test, we
substitute
\begin{equation}
x\left(  \tau\right)  =x_{0}\left(  \tau-\tau_{0}\right)  ^{-\frac{1}{2}%
}+x_{1}\left(  \tau-\tau_{0}\right)  ^{0}+x_{2}\left(  \tau-\tau_{0}\right)
^{\frac{1}{2}}+x_{3}\left(  \tau-\tau_{0}\right)  +...~, \label{ww.39}%
\end{equation}
into~ (\ref{ww.38A}) where we find
\[
x_{1}=\frac{x_{0}^{2}}{3\sqrt{6}}~,~x_{2}=\frac{x_{0}\left(  x_{0}%
^{2}+18\left(  \gamma-2\right)  \right)  }{72}~,~...~.
\]

We conclude that equation (\ref{ww.38A}) possesses the Painlev\'{e} property
and the analytic solution is described by the Puiseux expansion.

\subsection{Case $\Lambda\neq0$}

We focus now on the case for which there exists a nonzero cosmological
constant term. In order to determine the leading-order term for the system
(\ref{ww.36A}), (\ref{ww.37A}) we substitute%
\begin{align}
\Omega_{\Lambda}\left(  \tau\right)   &  =\Omega_{\Lambda0}\left(  \tau
-\tau_{0}\right)  ^{q}~,\\
x\left(  \tau\right)   &  =x_{0}\left(  \tau-\tau_{0}\right)  ^{p}\text{ }%
\end{align}
and as above we determine $p=-\frac{1}{2}$ and $p=-1$ with constraint equation
$\lambda\left(  \gamma-2\right)  x_{0}^{2}-1=\gamma\Omega_{\Lambda0}$. Easily
we see that for $\Omega_{\Lambda}=0$, the previous leading-order term is recovered.

For the resonances we replace
\begin{align}
\Omega_{\Lambda}\left(  \tau\right)   &  =\Omega_{\Lambda0}\left(  \tau
-\tau_{0}\right)  ^{-1}+n\left(  \tau-\tau_{0}\right)  ^{-1+S},\\
x\left(  \tau\right)   &  =x_{0}\left(  \tau-\tau_{0}\right)  ^{-\frac{1}{2}%
}+m\left(  \tau-t_{0}\right)  ^{-\frac{1}{2}+S}\text{ }%
\end{align}
in (\ref{ww.36A}), (\ref{ww.37A}).

We expand around $m^{2}\rightarrow0$,~$n^{2}\rightarrow0$ and $mn\rightarrow
0$. The first-order perturbed terms provide the matrix
\begin{equation}
A=%
\begin{pmatrix}
-\frac{2}{x_{0}}\left(  \Omega_{\Lambda0}\left(  \gamma\Omega_{\Lambda
0}-1\right)  \right)  & -\left(  S+\gamma\Omega_{\Lambda0}\right) \\
\left(  \left(  1+S\right)  -\gamma\Omega_{\Lambda0}\right)  & \frac{1}%
{2}x_{0}\gamma
\end{pmatrix}
,
\end{equation}
where the zeros of the determinant $\det\left(  A\right)  =0,~$give the
resonances; they are
\begin{equation}
S_{1}=-1,~S_{2}=0.
\end{equation}

The resonances indicate that the there exists a movable singularity and that
one of the coefficients, $x_{0}$ or $\Omega_{\Lambda0}$, is arbitrary and is
the second integration constant of the dynamical system.

We write the Puiseux series
\begin{equation}
\Omega_{\Lambda}\left(  \tau\right)  =\Omega_{\Lambda0}\left(  \tau-\tau
_{0}\right)  ^{-1}+\Omega_{\Lambda1}\left(  \tau-\tau_{0}\right)  ^{-\frac
{1}{2}}+\Omega_{\Lambda2}\left(  \tau-\tau_{0}\right)  ^{0}+\Omega_{\Lambda
3}\left(  \tau-\tau_{0}\right)  ^{\frac{1}{2}}+...~, \label{ww.40A}%
\end{equation}%
\begin{equation}
x\left(  \tau\right)  =x_{0}\left(  \tau-\tau_{0}\right)  ^{-\frac{1}{2}%
}+x_{1}\left(  \tau-\tau_{0}\right)  ^{0}+x_{2}\left(  \tau-\tau_{0}\right)
^{\frac{1}{2}}+x_{3}\left(  \tau-\tau_{0}\right)  +...~ \label{ww.41A}%
\end{equation}
which we replace in (\ref{ww.36A}), (\ref{ww.37A}). Hence, the latter Puiseux
series solves the dynamical system when%
\begin{equation}
\Omega_{\Lambda1}=\frac{4x_{1}\Omega_{\Lambda0}\left(  \gamma\Omega_{\Lambda
0}-1\right)  }{x_{0}\left(  1+2\gamma\Omega_{\Lambda0}\right)  },~x_{1}%
=\frac{x_{0}\left(  \sqrt{6}x_{0}\left(  1-2\Omega_{\Lambda0}\right)
+6\gamma\Omega_{\Lambda1}\left(  \gamma\Omega_{\Lambda0}-1\right)  \right)
}{6\left(  \gamma\Omega_{\Lambda0}-1\right)  \left(  2\gamma\Omega_{\Lambda
0}-3\right)  }~,~...~.~
\end{equation}

We summarize our results in the following proposition.

\textbf{Proposition 1: }\textit{The cosmological field equations in Weyl
Integrable theory for a spatially flat FLRW background geometry induced with a
cosmological constant term and an ideal gas form a dynamical system which
possesses the Painlev\'{e} property that is, the field equations are
integrable and the analytic solution is expressed by the Puiseux expansions
(\ref{ww.40A}) and (\ref{ww.41A}).}

\section{Conclusions}

\label{sec5}

In this piece of study we investigated the integrability properties of the
field equations in Weyl Integrable theory. In particular we studied the
existence of analytic solutions of the cosmological scenario for a universe
with matter source that of an ideal gas in a spatially flat FLRW geometry with
a nonzero cosmological constant term. For our analysis we applied the
singularity analysis which has been widely applied in gravitational physics
with many interesting results.

The field equations for the model of our consideration have the property to
admit a minisuperspace. That means that the field equations follow from the
variation for the dynamical variables of a point-like Lagrangian. For the case
in which there is no cosmological constant term, from previous results we know
that there exists a second conservation law which indicates the integrability
properties for the dynamical system. However, in the presence of the
cosmological constant, this conservation law does not exist. Thus we were not
able to infer about the integrability of the field equations.

For simplicity of our analysis we wrote the field equations into the
equivalent system with the use of dimensionless variables. For the equivalent
system we applied the singularity analysis and we found that the field
equations possess the Painlev\'{e} property, that is, the cosmological model
of our analysis is integrable with or without the cosmological constant term.
This is a very interesting result because after the loss of the additional
integration constant provided by Noether's analysis we were not able to make
an inference about the integrability and someone could infer that the
introduction of the cosmological constant into the field equations may violate
the integrability property.

As we have discussed before, this cosmological model describes important eras
of the cosmological history. The existence of the analytic solution indicates
that there exist an actual real solutions which correspond to the numerical
simulations of the field equations. This is an important feature because we
can understand the evolution of the dynamical system according to the free
parameters. In addition the above analysis can be used for the analytic
reconstruction of the cosmographic parameters \cite{com1,com2} and the
relation of these parameters with the initial value problem.

This work contributes to the subject of the derivation of exact and analytic
solutions in gravitational physics, while it shows that Noether symmetry
analysis and the singularity analysis are two complementary methods which can
provide interesting results.

\end{document}